# Frequency and time transfer for metrology and beyond using telecommunication network fibres

# Transfert d'une référence de temps ou de fréquence par fibre optique pour la métrologie et au-delà

Olivier Lopez[1], Fabien Kéfélian[1], Haifeng Jiang[2], Adil Haboucha[2], Anthony Bercy[1,2], Fabio Stefani[1,2], Bruno Chanteau[1], Amale Kanj[2], Daniele Rovera[2], Joseph Achkar[2], Christian Chardonnet[1], Paul-Eric Pottie[2], Anne Amy-Klein[1], Giorgio Santarelli[3*]

[1] Laboratoire de Physique des Lasers, Université Paris 13, Sorbonne Paris Cité, CNRS, 99 Avenue Jean-Baptiste Clément, 93430 Villetaneuse, France.
[2] LNE-SYRTE, Observatoire de Paris, CNRS, UPMC, 61 Avenue de l'Observatoire, 75014 Paris, France
[3] Laboratoire Photonique, Numérique et Nanosciences, UMR 5298, Institut d'Optique Graduate School, CNRS and Université de Bordeaux, 1, Rue F. Mitterrand, 33400 Talence, France
* Corresponding author : +33557017250, fax, giorgio.santarelli@institutoptique.fr

**Abstract**

The distribution and the comparison of an ultra-stable optical frequency and accurate time using optical fibres have been greatly improved in the last ten years. The frequency stability and accuracy of optical links surpass well-established methods using the global navigation satellite system and geostationary satellites. In this paper, we present a review of the methods and the results obtained. We show that public telecommunication network carrying Internet data can be used to compare and distribute ultra-stable metrological signals over long distances. This novel technique paves the way for the deployment of a national and continental ultra-stable metrological optical network.

**Résumé**

La distribution et la comparaison d'étalons de fréquence optique ultra-stables et d'échelle de temps ont été grandement améliorées depuis dix ans par l'emploi de fibres optiques. La stabilité de fréquence et l'exactitude des liens optiques fibrés surpassent les méthodes bien établies fondées sur les communications satellitaires. Dans cet article, nous présentons les méthodes et les résultats obtenus pendant cette décade. Nous montrons que les réseaux de télécommunication publics transportant des données internet peuvent être utilisés pour comparer et distribuer des signaux métrologiques sur de grandes distances. Ceci ouvre la voie au déploiement d'un réseau métrologique à l'échelle nationale et continentale.

1. **Introduction**

Frequency metrology has developed considerably over the past ten years and has benefitted from scientific advances in the areas of atomic laser cooling and femtosecond frequency combs. Today cold-atom microwave frequency standards routinely reach a fractional accuracy approaching $10^{-16}$ [1-3]. Trapped-ion or neutral-lattice optical clocks have already demonstrated accuracy in the low $10^{-17}$ and stability down to $10^{-18}$ or better in several laboratories [4-8]. This outstanding performance makes them ideal tools for laboratory tests of the validity of General Relativity (see for instance [7, 9, 10]). Among them, the comparison of different types of clocks is used to detect possible temporal variations of fundamental physical constants [11-13]. More generally, accurate time and frequency transfer is essential for relativistic geodesy, high-resolution radio astronomy, and is the basis of almost every type of precision measurement.

Until recently, the conventional means for remote frequency transfer was based on the processing of satellite radio-frequency signals, the Global Navigation Satellite System (GNSS), or dedicated satellite transfer [14]. Optical fibre links have shown potential to transfer frequency with much better accuracy and stability. The use of intensity modulated optical carriers around 1.55 μm (telecommunication window) to transfer radio-frequency or microwave signals have already been successfully demonstrated up to hundreds of kilometres over dedicated fibre routes [15-20]. A significant leap forward can be achieved using the optical phase of the optical carrier at 200 THz (1.55 μm) to transfer an extremely accurate and ultra-stable frequency reference over long distances. The high sensitivity detection of the optical phase using heterodyne methods in conjunction with spectrally very pure lasers, are the basic tools to achieve low-noise optical frequency transfer. In the last decade, several experiments in Europe, USA and Japan have explored the limits of this method, paving the way for a new generation of ultra-stable optical networks [18, 21-31]. A ground-breaking frequency transfer has been demonstrated on a record distance of 920 km and even 1840 km on a dedicated fibre network [27, 32]. The outcome of such a technique leads to high resolution comparisons between remote clocks in conjunction with the use of femtosecond frequency combs [26, 33-37].

In this paper, we will review nearly one decade of progress in optical carrier frequency transfer with optical fibres focusing on the use of public networks to achieve a scalable, continental-wide optical frequency distribution network with fibre links for science and industry.

2. **Optical carrier phase fibre link for frequency metrology: principle of operation**

We will focus now on optical carrier frequency transfer. The purpose of an optical link is to reproduce the frequency of an ultra-stable signal (seeded at the input end of the fibre) with the best fidelity at the output end. However, the optical phase is disturbed by the fluctuations of the optical path arising from the fibre temperature fluctuations and mechanical vibrations. To overcome this limitation, we need to measure and compensate for such fluctuations.

The first reported solution to this problem appeared in the mid-90s with the so-called "Doppler cancellation scheme" [38]. Figure 1 depicts the operation principle of a stabilized fibre link [39]. The light makes a round trip in the same fibre in order to experience twice the perturbations along the optical path. At the input end, one compares the input phase and the round-trip signal phase. The phase difference gives the fluctuations arising from the round-trip propagation in the fibre. It can be used to generate a correction signal at the input end, assuming that the forward single trip fluctuation is exactly half of the total round-trip fluctuations. Thus, this compensation set-up requires a high degree of symmetry between forward and backward paths since only reciprocal/symmetric phase fluctuations are corrected. It is worth noting that the link-delayed self-heterodyne phase noise of the laser source should be negligible compared to the free-running fiber phase noise [23]. This implies that this method also requires low-frequency-noise laser sources. For long links (i.e. beyond a few tens of kilometres), state-of-the-art cavity-stabilized lasers are therefore mandatory. Due to the propagation delay, $\tau$, along the fibre, the forward and backward noises are slightly different and the backward noise cannot catch up with the forward noise. This limitation was first highlighted by Newbury and co-workers [22, 40], who showed that the lowest achievable power spectral density (PSD) for optical phase noise is proportional to the PSD for free-running fibre propagation noise multiplied by $\tau^2$ and the Fourier frequency. Moreover, the correction bandwidth is severely reduced by the propagation delay, as in any time delayed control system. For a fibre link of 500 km, this bandwidth is below 100 Hz and the noise rejection slightly better than 40 dB at a Fourier frequency of 1 Hz. In principle, this can limit the optical link range especially for noisy links. In order to circumvent this potential problem and address other issue related to the fading of the signal along the fibre, our group introduced the concept of a cascaded link. In this approach, the propagation noise is compensated by successive fibre spans [23]. As the amplitude and the bandwidth of the noise correction are decreasingly linear with the length of the link, and assuming that the free-running fibre noise scales as the length of each subsection, one expects that the relative frequency stability of a cascaded link, divided into $N$ segments, is improved by a factor equal to $N^{1/2}$.

With optical links, signal attenuation due to the fibre losses and the numerous connectors and splices is also a major concern. To address this issue, one typically uses all-optical amplification techniques already demonstrated in public telecommunication networks. However, in our case the amplifiers must be bi-directional to allow the round-trip propagation on exactly the same optical path. The most widely used amplifier is the well-known Erbium-doped fibre amplifier. Brillouin [41, 42] and Raman amplifiers [43] have also been specifically developed for this application—enabling better signal amplification.

Excess phase noise in the measurement set-up can also limit the performance of the frequency transfer. The detection of phase noise in the link is performed by a Michelson-type interferometer with a short local reference arm, and the link acting as the second arm. The fibre noise arising from the local reference arm gives rise to so-called interferometric noise (mainly due to uncontrolled environmental fluctuations), which can dominate for long averaging times [44]. Additionally, the compensation system does not correct the non-reciprocal phase fluctuations between the forward and backward beams arising from, for instance, power imbalance or polarization effects such as polarisation mode dispersion (PMD). However, in the case of optical links, there was no evidence of these noise sources at a level of $10^{-19}$ or below. Among the non-reciprocal phase shifts, the Sagnac effect created by the Earth's rotation induces a phase offset which fluctuates mainly due to diurnal variations. The induced fractional-frequency instability is below $10^{-19}$ for a link up to a few thousand of kilometres at constant latitude of 45°[45].

## 3. Ultra-stable optical frequency transfer on long-haul fibre links

The first in-field long-distance optical carrier link was implemented in 2006 using a pair of 43-km telecommunication fibres connecting LNE-SYRTE and LPL in the Paris urban area, in collaboration with G. Grosche, B. Lipphardt and H. Schnatz from PTB [21]. Figure 2 displays the frequency instability obtained over this 86-km optical link (black filled triangles). It reaches $2 \times 10^{-15}$ for 1-s integration time and $6 \times 10^{-18}$ for $10^4$ s integration time. Similar results were obtained on a 76 km of installed fibres (251-km link using additional fibre spools of 175 km) by Newbury and co-workers at NIST [22]. This first demonstration of a highly-stable optical frequency transfer over fibres with active noise compensation has paved the way for a wide development of optical links all over the world. Following these experiments, our two groups of LPL and LNE-SYRTE have further investigated the 86-km link (see Fig. 2) and demonstrated a 172-km link using recirculation in the same fibres [23]. Following these pioneering results, several groups worldwide have initiated experiments on optical links [24, 27, 28, 30-32, 46-48].

A major challenge of the broad deployment of optical links is the difficulty (and the related high cost) to access dedicated fibre links between laboratories. In order to circumvent this problem, we explore an alternative approach of utilizing active-noise-compensated fibre links embedded in existing public telecommunication networks [49]. With this method, the ultra-stable optical signal shares the same fibre as the data traffic, using the widely deployed fibre network connecting every laboratory all over the world. Switching from a dedicated to a public fibre network is a major breakthrough for expanding ultra-stable frequency reference distribution to a larger scale in order to reach institutions and laboratories realizing precision measurements.

We will explain in detail this approach of ultra-stable optical frequency transfer, which is called the "dark channel" technique. We take advantage of the well-known and mature technology of Dense Wavelength Division Multiplexing (DWDM), where the fibre transmission spectral window is subdivided in a frequency grid with 100 GHz (0.8 nm) equally-spaced channels. One channel supports the propagation of the ultra-stable signal leaving the rest of the optical spectrum available for data transmission. In contrast to the so-called "dark fibre" approach, in this case the fibre is entirely dedicated to the propagation of the ultra-stable optical signal. While being very appealing, the dark channel technique imposes several conditions to be fulfilled. First of all, unlike the well-established long-haul telecommunication technique, the ultra-stable optical signal needs to propagate from one end of the link to the other along both directions, and in exactly the same fibre, such that the phase noise can be most accurately measured and compensated. This implies that presently operating unidirectional Erbium Doped Fibre Amplifiers (EDFA) must be bypassed, as well as any optical-to-electrical conversion stages (switches and routers). This can be accomplished by installing off-the-shelf passive optical filters, (Optical Add/Drop Multiplexers, OADM), able to inject or extract one single channel from the DWDM flux with very low insertion losses (about 1 dB per OADM). Additionally, custom-made fully bidirectional amplifiers need to be deployed—enabling forward and backward amplification. Figure 3 displays a scheme of such an optical link, between Paris and the city of Reims (see below).

We demonstrated for the first time such a multiplexed optical link at the end of 2008 using the French Academic and Research Network of RENATER [49]. This 108-km link was composed of two 43-km spans of dedicated fibres (connecting LNE-SYRTE and LPL), and a 22-km "dark channel" span that was deployed between LPL and a network node located at Aubervilliers. We obtained a relative frequency stability of $4\times10^{-16}$ at 1-s integration time and below $10^{-19}$ at $10^4$ s (comparable to that of the 86-km dedicated link), while leaving the data transmission unperturbed. This first significant result showed that the active noise compensation is compatible with a public telecommunication network opening the way for the development of an ultra-stable fibre network using the national RENATER network.

The natural follow-up to this work was to address long-haul optical links on a continental scale. This would allow the development of a fibre network in Europe connecting the national metrological institutes. When the link length exceeds a few hundred kilometres, the limited noise rejection amplitude, low control bandwidth, high fibre noise and low signal due to optical losses pose additional challenges compared to shorter link lengths. The latter issue cannot be entirely compensated with optical amplifiers for most links. Specifically, on the link from Paris to Strasbourg, due to back reflections and stray light from the public fibre network, we have found that the optical gain of the bi-directional EDFAs must be kept below about 15 dB to avoid self-oscillation.

To overcome this limitation, as mentioned before, we introduced the concept of cascaded links. With this approach, the optical signal is repeated from one span to the other using a so-called repeater station. Such a station contains a laser source to boost the optical signal injected in the successive span. This laser is phase-locked to the incoming signal (with a small frequency offset). Part of the light of the remote laser is sent backward to enable the stabilization of the upstream link. The other part of the light is sent forward and allows the stabilization of the following span. A small fraction of the station laser optical power can potentially be used to deliver a stable and accurate optical frequency to a user nearby.

A sketch of the repeater station is shown in Fig. 4. The apparatus is designed to operate autonomously outside the lab. Specifically, no stable RF oscillator is required for the transfer of the ultra-stable signal to the next fibre span [25]. At the input, the beat-note signal detected on the photodiode PD1 is used to phase-lock the local laser to the incoming signal. The amplitude of this beat note signal is automatically maximized with an electronically driven polarization controller. An acousto-optic modulator (AOM), used as a frequency shifter, compensates for the propagation noise of the following span as detected on the photodiode PD2. All of the electronic sub-systems and servo systems are driven by microcontrollers. An embedded computer can modify the control loop settings, supervise the system and communicate with IP protocols. The optical system of the regeneration laser station is carefully designed to minimize temperature sensitivity so that the phase fluctuations induced by temperature are nearly negligible. We have achieved an interferometer system with a very low temperature sensitivity of 1 fs/K [44].

With such a station, we demonstrated a 540-km cascaded link composed of two sub-links using two parallel fibres between Villetaneuse and Reims and back. The link starts and ends at the LPL laboratory (Université Paris 13) and the repeater station is located at Reims. Each sub-link is composed of three different fibre spans [28]. The first span of 11 km and the last span of 223 km are simultaneously carrying digital data traffic from the RENATER network. The last span consists of three subsections connected with EDFAs, as displayed on Fig. 3. The middle span is a dedicated dark fibre of 36 km in the Paris region. Eight OADMs are used for each sub-link to insert and extract the ultra-stable signal from the Internet fibres. The attenuation of a 270-km link is 80 dB due to fibre losses, OADMs and the large number of connectors. This is partially compensated using three bidirectional EDFAs with a total amplification gain of about 50 dB. The uncompensated attenuation exceeds 30 dB on each sub-link.

The cascaded link stability is obtained by frequency counting the end-to-end beat note signal using a dead-time-free counter with 1-s gate time and lambda-type operation. Here, lambda-type counting is used because it is more effective for noise filtering. For further details on counter operation, see Refs. [50, 51]. Figure 5 displays frequency stabilities computed from the recorded samples. The blue, upside-down triangles show the stability obtained from the lambda-type data using the modified Allan deviation. It is $10^{-16}$ at 1-s averaging time and reaches a noise floor of about $2\times10^{-20}$ after 4000 s. In the same figure, we also plot the free-running stability of the 540-km link (red squares), which shows a nearly flat value of $10^{-14}$ between 10 s and $10^4$ s. The green dotted line represents the lowest frequency stability reported to date for an optical clock [8], which is about 2 orders of magnitude above the optical link stability. This shows that this link can be used to transfer the frequency stability of the best optical clocks. We have also performed an evaluation of the accuracy of the frequency transfer. The mean value of the end-to-end beat note frequency offset is $-2.6\times10^{-20}$ with a conservative uncertainty of $1\times10^{-19}$ given by twice the 1-sigma upper limit of the long-term overlapping Allan deviation (black right-side-up triangles). We checked that we obtained the same long-term overlapping Allan deviation after averaging the data on 100 or 1000 successive points in order to eliminate any filtering effect associated with the lambda-type counting [50, 51]. The frequency stability and accuracy are compatible with single-span optical links using dedicated fibres [31, 32]. This is an unequivocal demonstration that public telecommunication networks can be used for ultra-stable and accurate frequency transfer. In conclusion, it also shows that state-of-the-art atomic frequency standards can be compared with negligible degradation in terms of both frequency stability and accuracy.

**4. Time Transfer through Optical Fibres (TTTOF)**

More recently, several groups have started experiments on highly-accurate time transfer over optical fibres. Here, the challenge is to control the one-way propagation delay with a high accuracy and a large dynamic. For this purpose, several methods are under development. One method relies on the stabilization of the propagation delay and a careful calibration procedure [18, 52]. Another is based on the well-known two-way methods, introduced in the 80's by Hanson [53]. We will focus on the latter, since we have demonstrated it on a fibre link spanning 540 km.

The method of two-way time transfer through optical fibres, as introduced in Ref. [54], is similar to the two-way method used in satellite time and frequency links [53]. At each end of the fibre link, a laser signal is sent to the other end and one detects the frequency difference between the local laser and the remote laser. Assuming that the propagation timing noise is equal for the two directions of propagation, one can efficiently reject the propagation contributions by synchronizing and post-processing the data (i.e. simply subtracting and dividing by two the two data sets recorded at each end). This method was first demonstrated on a few km-scale and then on a 73-km fibre link between Hannover and Braunschweig in Germany [55] with amplitude-modulated lasers. We extended it to the 540-km link between Villetaneuse and Reims, demonstrating simultaneously time comparison and ultra-stable frequency transfer [56]. Here, we use a single-span link without regeneration. The in-field implementation of our two-way method requires two ultra-stable lasers at each end, and an accurate control of their frequency drifts.

The experimental scheme is shown in Fig. 6. Time transfer signals are provided by a pair of two-way satellite time-transfer modems (TimeTech-SATRE). The principle of the time measurement is to relate the phase of a pseudorandom noise code (PRN), carried by a radio frequency carrier signal, to the one-pulse-per-second (1PPS) and a 10 MHz reference signal from a common clock. The PRN code rate was set to 20 Mchip/s, and the codes allocated to each modem were chosen to be orthogonal. The modem correlates the signal received from the remote end with a local replica of the expected signal. This gives the time of arrival of the received signal with respect to the local clock. Finally, the time-of-arrival datasets from the two modems are collected by a computer, which calculated the differential time delays. Phase modulation through an electro-optic modulator was then used to encode the PRN code onto the optical carrier [56].

The frequency transfer stability and the timing error were simultaneously measured and are plotted in Fig. 7. The frequency stability of the link reaches a resolution of $10^{-18}$ at 30 000 s averaging time, which is almost identical to the one reported in Ref. [28]. The timing stability (TDEV) is less than 20 ps for any measurement time up to one day.

The uncertainty budget of the experiment has been established by performing several tests and is detailed in Ref. [56]. This leads to a conservative estimate of the uncertainty budget of 250 ps, dominated by scarce phase jumps of about 50-80 ps. We were not able to univocally understand the origin of these phase jumps. Nevertheless, the system is quite robust: the peak-to-peak fluctuations are below 200 ps, while one-way fluctuations exceed 10 ns.

This method of accurate time transfer can be improved by developing new modems with much faster PRN codes, for instance. The method can be drastically simplified if one focuses only on time transfer, since an ultra-stable laser is not required. The extension to cascaded optical links requires the development of new type of repeater stations with accurate time calibration procedures.

## 5. Current developments on fibre links

The field of optical links is rapidly developing, and new ideas are emerging on link techniques and their applications. For instance, it was recently proposed to extend the two-way time transfer approach to frequency comparisons of stable optical references [57, 58]. The stability of the comparison then critically depends on the data-sampling synchronization accuracy. In this case, the set-up complexity is slightly reduced since no active compensation loop is required. An alternative set-up was also proposed by our group with two parallel fibres and a uni-directional propagation

[58]. The main motivation is to minimize the impact on the public telecommunication network and gain access to networks with underwater trans-oceanic spans. A proof-of-principle experiment was performed on a 100-km fibre link on demonstrated frequency comparisons with a relative frequency stability below $10^{-16}$ at 2000 s of integration time. The loss of reciprocity in the light path propagation, while being detrimental for metrological applications, is appealing for frequency comparisons of microwave frequency standards and atomic and molecular spectroscopy.

The fibre link community is also actively addressing the issue of distributing stable and accurate optical and radio-frequency references to many users. For this purpose, several groups are investigating new methods to extract from a stabilized optical link a reference signal with lower complexity [58-61].

Long-distance fibre links can also be used to consistently evaluate and calibrate the well-established GNSS or two-way satellite time and frequency transfer techniques thus supporting and improving these more classical methods. A first comparison with GPS was performed and reported in Refs. [62, 63].

Alternative approaches for precise and accurate time transfer are implemented using high speed 10Gb/s data transmission [64] and using ultra-stable optical carrier links with controlled frequency chirps [65].

In contrast to high-accuracy time and frequency transfer for advanced frequency metrology, lower accuracy approaches based on the Internet protocol like the White Rabbit or the PTP methods are now intensively studied for wide-area distribution with minimal network modifications and enhanced Internet hardware [66].

In addition, a strong impact of long distance optical links is expected for the assessment of the ACES space mission time and frequency transfer capability [67].

The deployment of optical fibre links can potentially impact geodesy and Earth sciences, for instance, by detecting the Sagnac phase fluctuations when a long-haul fibre link forms a closed loop—enclosing an extremely large area. A first theoretical treatment is given in Ref. [68], and a preliminary experimental loop of 47 km was reported in Italy [69]. Even if it seems difficult to achieve competitive rotation sensing compared to the best laser gyrometer [70], the potential of Sagnac detection with giant fibre loops is still an interesting option, because the measurement is no longer localized to a small portion of space. In addition, optical fibre links can be used to perform relativistic geodesy by measuring the frequency of highly-reproducible, transportable optical clocks that are sensitive to the gravitational potential of the Earth [45, 71].

Beyond the frequency and time metrology community, the wide deployment of optical links, frequency combs and ultra-stable optical cavities can provide stable and accurate optical references to research labs. This will open a new window for high resolution spectroscopy and precise measurements such as, for example, parity non-conservation [72] or Hydrogen spectroscopy [34].

6. Conclusion and outcome

After nearly one decade of development, we have shown that the field of optical fibre frequency transfer is now mature enough to support ultra-stable optical signal distribution and accurate and reliable clock comparisons over long distances. Several projects are aiming at the development of national fibre networks for metrological optical signals in Europe and in Asia. The French project named REFIMEVE+ coordinated by our two groups is developing a wide national infrastructure where a reference optical signal generated at SYRTE Paris Observatory is distributed to about 20 academic and institutional users using the RENATER network. In addition, this project targets transnational links towards Germany, the UK and Italy as well as a giant fibre loop with a length of a few thousand kilometres. The deployment of such a network is expected to start at the end of 2016, while preliminary branches are presently under development to realize international clock comparisons at the European level. In Europe there are several initiatives to establish national ultra-stable networks for instance in Italy with the LIFT project and in Germany with the Braunschweig-Munchen links. Several aspects of time and frequency transfer with optical fibres are also investigated in a number of European countries and throughout the European consortium NEAT-FT.

One challenging target is to perform very-high resolution clock measurements on accurate optical clocks that are being developed in France, Germany and the UK. For this purpose, an optical fibre link connecting Paris (SYRTE) to Braunschweig (PTB) is now complete with promising preliminary results, and a connection through the English Channel underwater tunnel is presently in progress. We are confident that in less than one year we will be able to compare optical clocks at the state-of-the-art levels using fibre links. This will strengthen frequency metrology, and enable temporal stability tests of fundamental constants by comparing distant uncorrelated optical clocks using different atomic species. The deployment of an ultra-stable fibre network will also pave the way for the use of optical clocks to perform relativistic geodesy, and will enable precision experiments in any laboratory connected to a metrological institute. We anticipate that the development of ultra-stable frequency distribution will stimulate the development of a broad range of new applications.

Acknowledgments

We acknowledge the financial support from the Agence Nationale de la Recherche (ANR BLANC 06-3_144016 and 2011-BS04-009-01), Labex First-TF, the European Metrology Research Programme (contract SIB-02 NEAT-FT), IFRAF-Conseil Régional Ile-de-France, the French space agency CNES, GRAM, and the continuous support of RENATER. The authors especially thank Emilie Camisard.

**Figure and table caption**

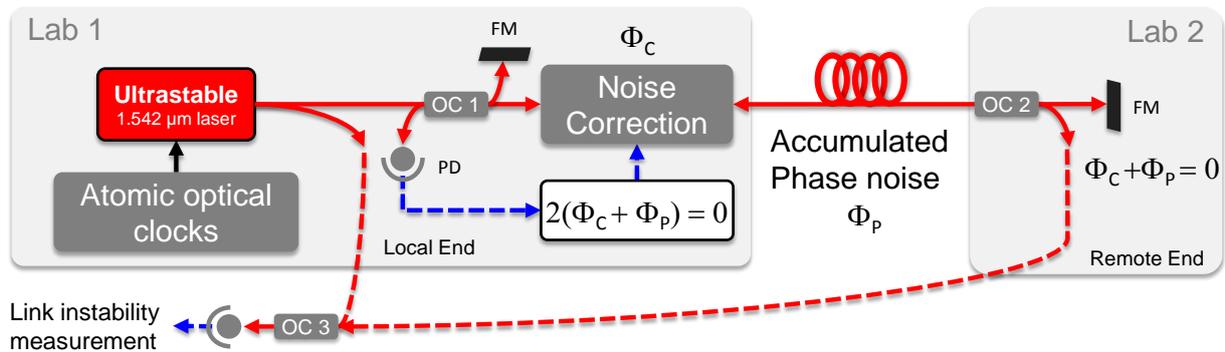

Figure 1: principle of an active-noise-compensated link. The propagation phase noise is detected using the round-trip signal and an active correction is implemented at the link input. The end-to-end beat note signal is detected to evaluate the link stability and accuracy. FM: Faraday mirror, OC: optical coupler.

Figure 1 : Principe de la compensation active du bruit d'un lien optique. Le bruit de propagation est détecté grâce à l'aller-retour du signal dans la fibre. La correction est effectuée au départ du lien. Le signal de battement entre les deux extrémités du lien permet d'évaluer la stabilité et l'exactitude du transfert de fréquence. FM : miroir de Faraday, OC : coupleur optique.

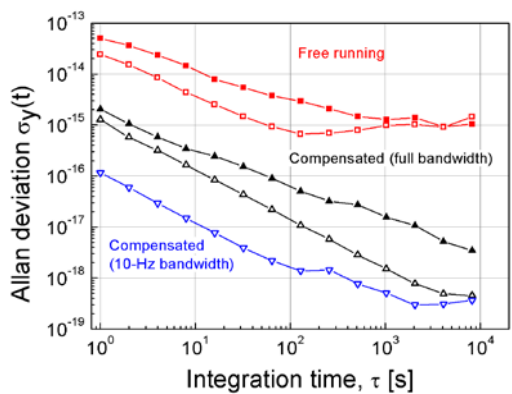

Figure 2: Fractional frequency instability versus averaging time of the LPL-SYRTE 86-km optical link for two experiments: a) first data recorded by the French-German collaboration in November 2006 [21] with a free-running link (red filled squares) and a compensated link with full bandwidth (black filled triangles); b) second experiment completed at the end of 2007 [23] with a free-running link (red open squares), a compensated-link with full bandwidth (black open triangles), and a compensated link with a 10-Hz measurement bandwidth (blue open upside-down triangles). Data were recorded with a dead-time-free counter in Π-mode (error bars are omitted for clarity).

Figure 2 : Instabilité relative de fréquence en fonction du temps de mesure, pour le lien LPL-SYRTE de 86 km, lors de deux expériences : a) premières mesures obtenues en novembre 2006 lors d'une expérience réalisée en commun par les groupes allemands et français [21], lien libre

(carrés pleins rouges) et lien compensé sans filtrage du signal (triangles pleins noirs), b) deuxièmes mesures réalisées fin 2007 [23], lien libre (carrés rouges), lien compensé sans filtrage du signal (triangles noirs) et lien compensé avec filtrage dans une bande de 10 Hz (triangles renversés bleus). Les données ont été enregistrées avec un compteur sans temps mort en mode Π et les barres d'erreurs ont été omises pour faciliter la lecture.

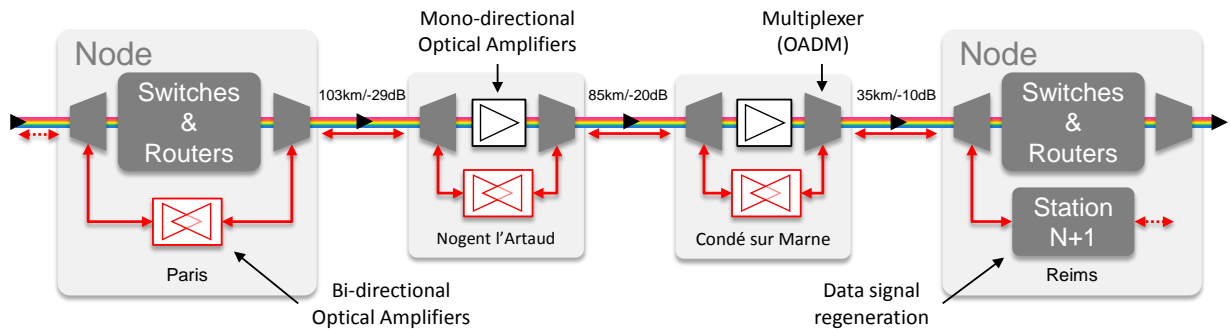

Figure 3: Scheme of an optical link between Paris and Reims using the RENATER network, with optical losses.

Figure 3 : Schéma d'un lien Paris-Reims sur le réseau RENATER, avec les valeurs des pertes optiques.

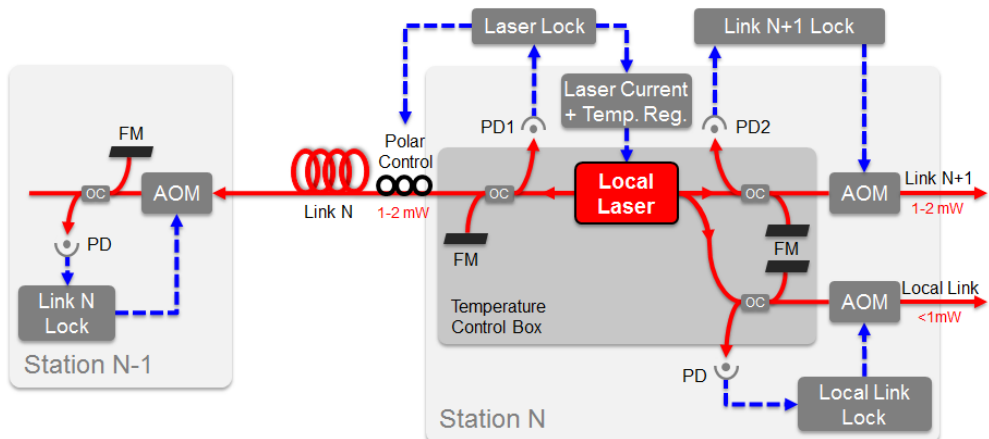

Figure 4: Schematic of the $N^{th}$ repeater station, FM: Faraday mirror, PD: photodiode, PC: polarization controller, AOM: acousto-optic modulator.

Figure 4 : Schéma de la $N^{ème}$ station répétitrice, FM : miroir de Faraday, PD : photodiode, PC : contrôleur de polarisation, AOM : modulateur acousto-optique.

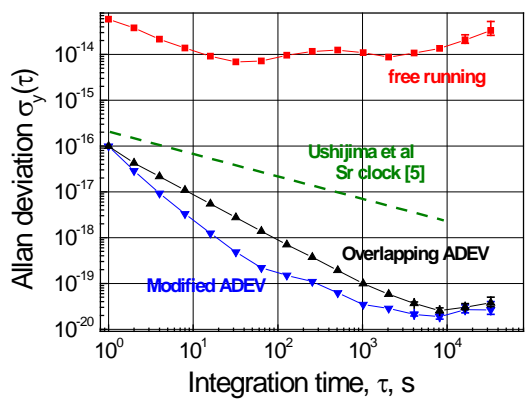

Figure 5: Fractional frequency instability versus averaging time of the Villetaneuse-Reims-Villetaneuse 540-km optical link, with a free-running link (red squares), and a cascaded-compensated link calculated from Λ-type data using overlapping Allan deviation (black right-side-up triangles) and modified Allan deviation (blue upside-down triangles). The latter stability enables one to see the noise floor limitation.

Figure 5: Instabilité relative de fréquence en fonction du temps de mesure, pour le lien Villetaneuse-Reims-Villetaneuse de 540 km : lien libre (carrés rouges), lien compensé cascadé, déviation d'Allan avec recouvrement (triangles noirs) et modifiée (triangles bleus pointe en bas) calculée à partir des données enregistrées avec un compteur Λ. Cette dernière stabilité permet de mettre en évidence le plancher de bruit de la mesure.

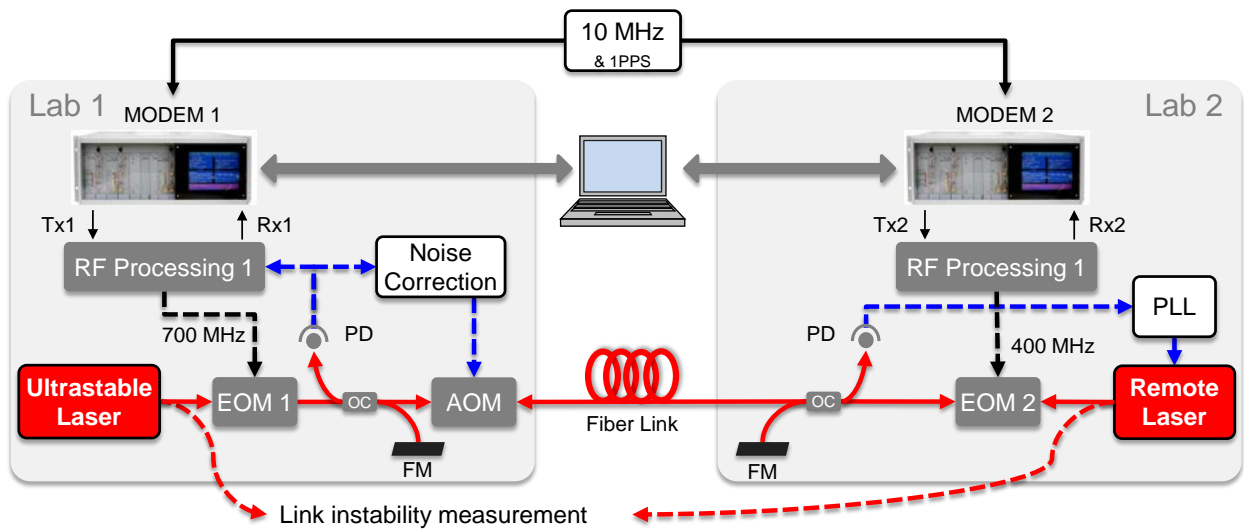

Figure 6: Scheme of the experimental set-up for simultaneous time comparison and frequency transfer; FM: Faraday mirror, PD: photodiode, OC: optical coupler, AOM: acousto-optic modulator, EOM: electro-optic modulator, PLL: phase-locked loop.

Figure 6 : Schéma du dispositif expérimental pour le transfert simultané de temps et de fréquence; FM : miroir de Faraday, PD : photodiode, OC : coupleur optique, AOM : modulateur acousto-optique, EOM : modulateur électro-optique, PLL : boucle d'asservissement en phase.

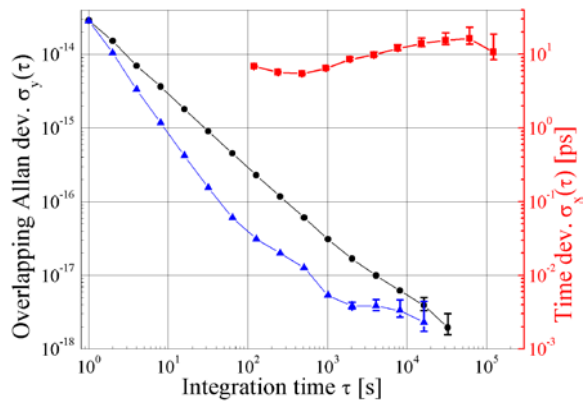

Figure 7: Time-transfer stability for the 540-km fibre link. a) Fractional optical-frequency stability calculated as the overlapping Allan deviation (black circles, left scale); b) the modified Allan deviation (blue triangles, left scale); c) timing stability in picoseconds (red squares, right scale). Data (full bandwidth) were recorded with a dead-time-free counter in Π-mode.

Figure 7 : Instabilité relative du transfert de temps pour le lien de 540 km, a) déviation d'Allan avec recouvrement (ronds noirs, échelle de gauche) b) déviation d'Allan modifiée (triangles bleus, échelle de gauche), c) déviation en temps en ps (carrés rouges, échelle de droite). Les données sans filtrage) ont été enregistrées avec un compteur sans temps mort en mode Π.